\documentclass[]{jfm}
\usepackage{bm}
\usepackage{graphicx}
\usepackage{newtxtext}
\usepackage{newtxmath}
\usepackage{natbib}
\usepackage{hyperref}
\hypersetup{
    bookmarks  = false,
    colorlinks = true,
    urlcolor   = blue,
    citecolor  = black,
}
\hypersetup{pdfstartview={XYZ null null null}}

\newcommand{\RomanNumeralCaps}[1]\linenumbers

\newcommand{\figpath}{}
\newcommand{\pfrac}[2]{\frac{\partial #1}{\partial #2}}
\newcommand{\e}{\mathrm e}
\newcommand{\ii}{\mathrm i}
\newcommand{\hy}{{\bar h}}
\DeclareMathOperator{\sech}{sech}

\title{Two-interface and thin filament approximation in Hele--Shaw channel flow}

\author{Michael C. Dallaston\aff{1}
  \corresp{\email{michael.dallaston@qut.edu.au}},
  Michael J. W. Jackson\aff{1},
  Liam C. Morrow\aff{2}
 \and \\ Scott W. McCue\aff{1}}

\affiliation{\aff{1}School of Mathematical Sciences, Queensland University of Technology, Brisbane QLD, 4001, AUSTRALIA
\aff{2}Department of Engineering Science, University of Oxford, Oxford OX1\,3PJ, UK}

\begin{document}
\maketitle

\begin{abstract}
When a viscous fluid partially fills a Hele--Shaw channel, and is pushed by a pressure difference, the fluid interface is unstable due to the Saffman--Taylor instability.  We consider the evolution of a fluid region of finite extent, bounded between two interfaces, in the limit the interfaces are close, that is, when the fluid region is a thin liquid filament separating two gases of different pressure.  In this limit, we derive a second-order `thin filament' model that describes the normal velocity of the filament centreline, and evolution of the filament thickness, as functions of the thickness, centreline curvature and their derivatives.  We show that the second-order terms in this model, that include the effect of transverse flow along the filament, are necessary to regularise the instability.  Numerical simulation of the thin filament model is shown to be in accordance with level-set computations of the complete two-interface model.  Solutions ultimately evolve to form a bubble of rapidly increasing radius and decreasing thickness.
\end{abstract}


\section{Introduction}
\label{sec:intro}

For a standard model of Hele--Shaw flow in a rectilinear channel, consisting of a semi-infinite inviscid fluid region and a semi-infinite viscous fluid region separated by a single interface, the interface exhibits the Saffman--Taylor instability when the inviscid fluid displaces the viscous, and is stable when the viscous fluid displaces the inviscid~\citep{Saffman1958}.  In the absence of surface tension, exact solution methods exist that exhibit either finite time singularity or long-time finger formation \citep{Howison1986,Howison1986a}; however, the problem is ill-posed.  A regularisation such as surface tension is needed to stabilise sufficiently large wavenumber perturbations.  In the presence of surface tension, solutions generically tend to a linearly stable travelling wave solution known as the Saffman--Taylor finger, although for very small surface tension, finite-amplitude noise in experiments and numerical simulations can lead to tip-splitting \citep{Casademunt2004}.

The traditional Saffman--Taylor instability is normally studied on the assumption that the viscous fluid region extends infinitely far along the channel, so that there is only a single interface to consider.  However, in reality the fluid region will only have finite extent, so that there are in fact two interfaces: one in which the viscous fluid is being displaced by the driving inviscid fluid, and one in which the viscous fluid is the one advancing (see Fig.~\ref{fig:schematic}).  In this case, the force driving the fluid region is the pressure difference between the two inviscid fluids on either end of the fluid region.  In the absence of surface tension, some classes of exact solutions to the two-interface problem have been found through use of special functions, in both channel geometries \citep{Crowdy2004,Feigenbaum2001} as well as annular geometries \citep{Crowdy2002,Dallaston2012}.  The exact solutions in these studies, however, do not exhibit the interfaces becoming closely separated.  \citet{Richardson1982,Richardson1996} similarly finds exact solutions for two-interface channel flow, some of which exhibit topological changes when two cusps form on different parts of the interface at the same point in space (none of these solutions correspond to the two semi-infinite regions of different pressure meeting, however).
\citet{Farmer2006} consider an approximate model where the two interfaces in an unbounded Hele--Shaw cell are very close, resulting in a thin filament of viscous liquid, and construct exact solutions to this approximate model (see Section \ref{sec:unregularized}).    All of these zero-surface-tension models are ill-posed.

The two-interface Hele--Shaw model in the limit that the interfaces are close is of great importance as, even if the fluid region is not initially thin, the effect of the Saffman--Taylor instability will result in a thin fluid region or filament developing (see for example the experimental results in \citet{Ward2011b,Morrow2023}, and \cite{Cardoso1995}, although the latter study focuses on a much less viscous thin annular region in a more viscous ambient fluid).  This formation of a thin filament precedes the fluid `bursting', at which point the two inviscid regions meet and the pressure rapidly equalises.  The breakup of a thin viscous filament in a Hele--Shaw cell (with surface tension but in the absence of a driving pressure difference) has also been examined by the use of the lubrication approximation \citep{Almgren1996, Almgren1996b,Constantin1993,Dupont1993,Goldstein1993,Goldstein1995,Goldstein1998}.  The complicated, but self-similar break-up behaviour of the filament in particular (where the filament thickness goes to zero at a finite time and point in space) is detailed in \citet{Almgren1996b}.

In this article we consider two-interface Hele--Shaw flow in a channel, including the effects of both driving pressure difference and surface tension, with particular focus on the case in which the fluid interfaces become close together.   In Section \ref{sec:formulation} we describe the full two-interface model and its stability.  In Section \ref{sec:thinfilament} we derive a simplified model (the thin filament approximation) that applies when the two interfaces are close together, by applying the lubrication approximation (up to second order) in a coordinate system following the filament centreline.  The application of lubrication theory in an evolving, curvlinear coordinate system is similar to that applied in \citet{Vandefliert1995} and \citet{Howell2003}, although taking such an approximation to higher order is not standard.  Our approximation also represents a generalisation of the model by \citet{Farmer2006} by including both the effects of surface tension, as well as higher order terms, which include the effect of flow along the filament.  In section \ref{sec:leadingOrder} we demonstrate the importance of including these higher order terms by showing (both numerically and by constructing similarity solutions) that the leading order problem generically blows up in finite time, even in the presence of surface tension.  We also find quasi-travelling wave solutions, which may be thought of as the analogue of Saffman--Taylor fingers, which generalise the `grim reaper' exact solution found by \citet{Farmer2006}.  
In Section \ref{sec:numerics} we compute numerical solutions of the thin filament model including the higher order regularising terms.  Solutions of the original two-interface model, found using a level set method, are seen to approach the thin filament model results as the resolution of the level set method is increased.  We show the general behaviour of our thin filament model is not to tend toward a quasi-travelling wave solution, but instead develop a rapidly expanding `bubble' of circular shape and decreasing thickness.

\section{Formulation of the two-interface Hele--Shaw flow problem}
\label{sec:formulation}

\subsection{Hele--Shaw flow equations}

\begin{figure}
\centering
\includegraphics{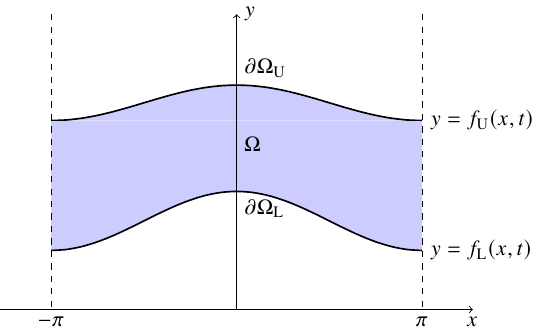}
\caption{A finite fluid region in a Hele--Shaw channel.  The fluid is driven in the positive $y$ direction by a pressure difference.}
\label{fig:schematic}
\end{figure}

We consider flow in a Hele--Shaw channel of nondimensional width $2\pi$ in the $x$-direction.  The fluid region $\Omega$ is bounded above and below by interfaces $\partial\Omega_\mathrm{U}$ and $\partial\Omega_\mathrm{L}$, respectively. A nondimensional pressure difference $P$ acts to push the fluid region in the positive $y$-direction, while surface tension acts on both interfaces (see Fig.~\ref{fig:schematic}).  The standard governing equations in nondimensional form are \citep[e.g.]{Morrow2021}
\begin{subequations}
\label{eqs:fullproblem}
\begin{align}
\nabla^2\phi = 0, \qquad &\bm x \in \Omega \label{eq:Laplace}\\
v_\mathrm{n}  = \pfrac{\phi}{n},  \qquad &\bm x \in \partial\Omega_\mathrm{L}, \partial\Omega_\mathrm{U} \label{eq:kinematicBCs} \\
\phi = -P - \sigma\kappa_\mathrm{L}, \qquad & \bm x \in \partial\Omega_\mathrm{L} \label{eq:pressureBCL}\\
\phi = \sigma\kappa_\mathrm{U}, \qquad & \bm x \in \partial\Omega_\mathrm{U} \label{eq:pressureBCU}
\end{align}
\end{subequations}
with $\phi$ the velocity potential, $v_\mathrm{n}$ the normal velocity of each interface, $\kappa_\mathrm{U}$ and $\kappa_\mathrm{L}$ the curvatures of each interface, $P$ is the nondimensional pressure difference, and $\sigma$ the nondimensional surface tension.
In \eqref{eqs:fullproblem} the normal vector $\hat{\bm n}$ of each interface is defined so that a flat interface has normal in the positive $y$-direction, and the signs of curvatures $\kappa$ are chosen such that positive $\kappa$ implies a concave interface in the positive $y$-direction (for example, both interfaces in Fig.~\ref{fig:schematic} have negative curvature at $x=0$ and positive curvature at $x=\pm\pi$). This choice of sign convention for both interfaces will simplify the thin filament derivation in the next section.  

For definiteness, we consider no-flux boundary conditions on the channel walls:
\begin{equation}
\pfrac{\phi}{x} = 0, \qquad y = \pm \pi,
\end{equation}
although many of our results, in particular the thin filament model derived in Section \ref{sec:thinfilament}, do not rely on these conditions, and could equally apply to an infinitely long fluid region in an unbounded cell, or an annular fluid region represented by a closed curve (in that case, the constant pressure difference $P$ would physically require injection or extraction of air into the interior of the cell).

The system \eqref{eqs:fullproblem} is derived from the dimensional system by introducing the length scale $[x]$, the pressure scale $[p]$, and defining the dimensionless velocity potential $\phi$ in terms of the pressure $p$:
\[
[x] = \frac{\ell}{2\pi}, \qquad [p] = \frac{12\mu [x]^2}{b^2[t]}, \qquad \phi = \frac{p_U-p}{[p]},
\]
so that
\[
P = \frac{p_U-p_L}{[p]} \qquad \sigma = \frac{\hat\gamma}{[p][x]},
\]
where $p_U$ and $p_L$ are the upper and lower inviscid fluid pressures, $b$ is the plate separation, $\ell$ is the channel width, $\mu$ the viscosity, $\hat\gamma$ is the surface tension, and $[t]$ is an arbitrary time scale.  While one of the parameters $P$ or $\sigma$ (if nonzero) could be scaled out by choosing an appropriate time scale, retaining both parameters aids in understanding the effects of the terms that arise in our later analysis.

\subsection{Linear stability}
\label{sec:fullLinearStability}
The two-interface Hele--Shaw configuration has an exact base state where both interfaces are horizontal, and the fluid region moves upward at constant velocity $v_\mathrm{n} = P/h_0$, where $h_0$ is the distance between the two interfaces.  To examine the stability of this configuration, write the system \eqref{eqs:fullproblem} in Cartesian ($x,y$) coordinates, and define the upper and lower interfaces as $y = f_\mathrm{U}(x,t)$ and $y = f_\mathrm{L}(x,t)$, respectively.  In addition to Laplace's equation for $\phi$, the boundary conditions (\ref{eq:kinematicBCs})-(\ref{eq:pressureBCU}) are
\begin{subequations}
\label{eqs:cartesianBCs}
\begin{align}
\phi|_{y=f_\mathrm{L}} = -P - \sigma \frac{(f_\mathrm{L})_{xx}}{(1+(f_\mathrm{L})_x^2)^{3/2}},  \qquad \phi|_{y=f_\mathrm{U}} = \sigma \frac{(f_\mathrm{U})_{xx}}{(1+(f_\mathrm{U})_x^2)^{3/2}}, \\
(f_\mathrm{L})_t = \phi_y|_{y=f_\mathrm{L}} - \phi_x|_{y=f_\mathrm{L}}(f_\mathrm{L})_x, \qquad (f_\mathrm{U})_t = \phi_y|_{y=f_\mathrm{U}} - \phi_x|_{y=f_\mathrm{U}}(f_\mathrm{U})_x
\end{align}
\end{subequations}
(for notational compactness we will use variable subscripts to refer to partial derivatives throughout this article; text or numerical subscripts, however, do not refer to partial derivatives).  The base state is (up to arbitrary translation in $y$) represented by $f_\mathrm{L} = (P/h_0)t - h_0$, $f_\mathrm{U} = (P/h_0)t$, and $\phi = (P/h_0) \hat y$, where $\hat y = (y-Pt/h_0)$ is the coordinate in the travelling frame.

\begin{figure}
\centering
\includegraphics{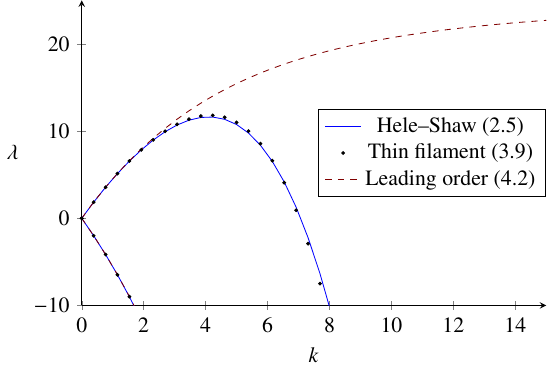}
\caption{Growth rate $\lambda$ of perturbations of given wavenumber $k$ for the two interface Hele--Shaw model \eqref{eq:full_eigs}, the filament model \eqref{eq:filament_eigs} (second-order in the lubrication parameter), and the leading-order version of this model, which lacks terms corresponding to transverse flow \eqref{eq:notransverse_eigs}.  Each model has the parameter values $P=1$, $h_0 = 0.2$, and $\sigma = 0.1$.  All systems have two eigenvalues for each wavenumber.  For the Hele--Shaw model and the filament model, the most unstable eigenvalue is stabilised at a cut-off wavenumber $k$, and agree closely.  The leading order filament model \eqref{eq:notransverse_eigs} is unstable for all wavenumbers, with the positive eigenvalue tending to a constant as $k\to\infty$.}
\label{fig:stability}
\end{figure}

To examine linear stability then, we impose perturbations on $f_\mathrm{L}$, $f_\mathrm{U}$:
\begin{align*}
f_\mathrm{L}(x,t) &= \frac{P}{h_0}t - h_0 + f_{\mathrm{L}1}(x,t), \\
f_\mathrm{U}(x,t) &= \frac{P}{h_0}t + f_{\mathrm{U}1}(x,t) \\
\phi(x,y,t) &= \frac{P}{h_0}\hat y + \phi_1(x,\hat y,t).
\end{align*}
On substitution into the boundary conditions,
\begin{align*}
f_{\mathrm{L}1t} &= \phi_{1y}, \qquad \phi_1 = -\sigma f_{\mathrm{L}1xx} - \frac{P}{h_0}f_{\mathrm{L}1}, \qquad \hat y = -h_0
\\
f_{\mathrm{U}1t} &= \phi_{1y}, \qquad \phi_1 = \sigma f_{\mathrm{U}1xx} - \frac{P}{h_0}f_{\mathrm{U}1}, \qquad \hat y = 0
\end{align*}
A perturbation with wavenumber $k$ in the $x$-direction takes the form
\[
f_{\mathrm{U}1} = A\cos(kx)\e^{\lambda t}, \qquad f_{\mathrm{L}1} = B\cos(kx)\e^{\lambda t}, \qquad \phi_1 = (c_1\e^{k\hat y} + c_2\e^{-k\hat y})\cos(kx)\e^{\lambda t},
\]
where the pressure boundary conditions give $c_1$ and $c_2$ in terms of the interfacial amplitudes $A$ and $B$, and the kinematic conditions result in an eigenvalue problem for $\lambda$:
\begin{equation}
\label{eq:fulleigenvalueproblem}
\lambda \begin{bmatrix} A \\ B \end{bmatrix} =
\frac{k}{\sinh(kh_0)} \begin{bmatrix} -\cosh(kh_0)(\sigma k^2 + P/h_0) & (-\sigma k^2 + P/h_0) \\ -(\sigma k^2 + P/h_0) & \cosh(kh_0)(-\sigma k^2 + P/h_0) \end{bmatrix} \begin{bmatrix} A \\ B \end{bmatrix}.
\end{equation}
The eigenvalues of this system are thus
\begin{equation}
\label{eq:full_eigs}
\lambda = -\sigma k^3 \coth(kh_0) \pm k\sqrt{\left(\frac{P}{h_0}\right)^2 + \frac{\sigma^2 k^4}{\sinh^2(kh_0)}}.
\end{equation}
The growth rate curves ($\lambda = \lambda(k)$) are shown in Fig.~\ref{fig:stability} (solid line) for parameter values $\sigma = 0.1, h_0 = 0.2$.  One eigenvalue is negative for all wavenumbers $k$, while the other is positive for a finite band of wavenumbers that ranges from zero up to a critical wavenumber.  The presence of surface tension regularises the system by stabilizing the wavenumbers for large $k$.  We emphasise that the system is unstable no matter the direction of the pressure gradient (regardless of the sign of $P$), as each direction involves one of the interfaces moving in the unstable direction according to the Saffman--Taylor instability.

Our linear stability analysis here for a finite viscous fluid evolving in a Hele--Shaw channel is analogous to that presented for a radial geometry with two interfaces \citep{Morrow2023}.  Generalisations and alterations to include different viscous fluids on either side of the interfaces have been conducted for both linear and weakly nonlinear frameworks \citep{Gin2015a,Anjos2020}.

We close this section by noting some limiting behaviour of (\ref{eq:full_eigs}). For large $h_0$, in order to keep the speed of the base state $O(1)$, we need to keep the driving pressure difference $P=O(h_0)$.  In that case, $\lambda\sim -\sigma k^3\pm kP/h_0$ as $h_0\rightarrow\infty$.  This limit agrees with the well-studied single interface problem (an infinite body of viscous fluid), with the plus (minus) sign associated with the unstable (stable) direction of flow.  Of a more particular interest here is the other limit of $h_0\ll 1$.  Again, supposing that $P=O(h_0)$ in order to keep the interface speed $O(1)$, we have
\begin{equation}
\label{eq:approx1}
\lambda \sim \left[-\sigma\left(\frac{k}{h_0}\right)^2\pm\left(\frac{k}{h_0}\right)
\sqrt{\left(\frac{P}{h_0}\right)^2 + \sigma^2\left(\frac{k}{h_0}\right)^2}
\right]h_0
\quad\mbox{as}\quad h_0\rightarrow 0, \quad k=O(h_0),
\end{equation}
\begin{equation}
\label{eq:approx2}
\lambda \sim \left\{\left[
\frac{1}{2\sigma}\left(\frac{P}{h_0}\right)^2-\frac{\sigma k^4}{2}
\right]h_0, \,
-2\sigma\left(\frac{k}{h_0}\right)^2\,h_0\right\}
\quad\mbox{as}\quad h_0\rightarrow 0, \quad k=O(1),
\end{equation}
\begin{equation}
\label{eq:approx3}
\lambda \sim -
\frac{\sigma(\cosh(kh_0)\mp 1)}{\sinh(kh_0)}
\frac{1}{h_0^3}
\quad\mbox{as}\quad h_0\rightarrow 0, \quad k=O(1/h_0),
\end{equation}
where for (\ref{eq:approx2}) $k=O(1)$ means strictly of order one, while for (\ref{eq:approx3}) $k=O(1/h_0)$ means strictly of order $1/h_0$.

\section{Thin filament approximation}
\label{sec:thinfilament}

\subsection{Model in intrinsic coordinate system}
In this section we derive an approximation of the two-interface flow by considering the thickness of the fluid region (that is, the distance between the two interfaces) to be small.  This approximation will be valid when the separation of the two interfaces is smaller than the radius of curvature of either interface, but still larger than the plate separation (if the interface separation is the same order as the plate separation, the depth averaging that leads to two-dimensional Hele--Shaw flow \eqref{eqs:fullproblem} is no longer valid).

\begin{figure}
\centering
\includegraphics{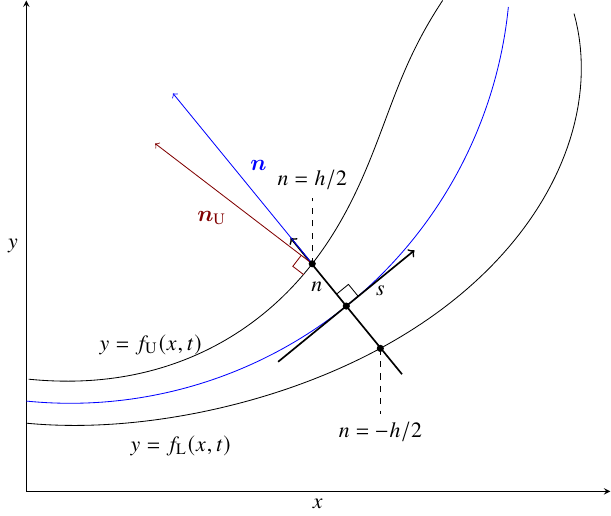}
\caption{A schematic of the coordinate system \eqref{eq:intrinsicCoordinates} used to derive the second-order lubrication model \eqref{eqs:filamentEqs}, which describe the velocity $v_\mathrm{n}$ of the centreline and the evolution of the film thickness $h$, in a direction normal to the centreline.  In the derivation of the model, it is important to distinguish between the centreline normal $\bm n$ and normals to the interface (e.g., $\bm n_{\mathrm{U}}$ on the upper interface), and velocity components in these respective directions.}
\label{fig:filamentSchematic}
\end{figure}

We will start by converting the governing equations and boundary conditions into an intrinsic coordinate system $(x,y) \mapsto (s,n)$ in which the filament is represented by a centreline curve $\bar{\bm x}(s,t) = (\bar x, \bar y)$, with $s$ the arclength coordinate, and $n$ the coordinate normal to this centreline; thus
\begin{equation}
\bm x = \bar{\bm x} + n\bm n
\label{eq:intrinsicCoordinates}
\end{equation}
\citep[e.g.,][]{Vandefliert1995, Howell2003}.  The coordinate system is represented schematically in Fig.~\ref{fig:filamentSchematic}.  Here $\bm n = (-\bar y_s, \bar x_s)$ is the centreline normal (again we use the variable subscript $s$ to represent the $s$ partial derivative, for brevity).  In this coordinate system we specify the upper and lower interfaces to occur at $n = \pm h(s,t)/2$, respectively, so that the normal thickness is $h$.  The upper interface is thus given by the curve
\[
\bm x_{\mathrm{U}} = \bm \bar {\bm x} + \frac{h}{2} \bm n,
\]
and a (non-unit) normal to this interface (as opposed to the centreline) is then
\[
\bm n_{\mathrm{U}} = \left(1 - \frac{h\kappa}{2}\right) \bm n - \frac{h_s}{2}\bm s,
\]
where $\bm s = (\bar x_s, \bar y_s)$ is the centreline unit tangent vector, and $\kappa = \bar x_s \bar y_{ss} - \bar y_s \bar x_{ss}$ is the centreline curvature.  

The velocity normal to the interface (not the centreline ) $v_{\mathrm{U}}$ is equal to the normal derivative of the potential via the kinematic condition \eqref{eq:kinematicBCs}:
\begin{align*}
v_{\mathrm{U}} &= \nabla \phi \cdot \frac{\bm n_{\mathrm{U}}}{|\bm n_{\mathrm{U}}|} \\
&= \frac{1}{|\bm n_{\mathrm{U}}|}\left[\pfrac{\phi}{n}\bm n + \frac{1}{1-n\kappa}\pfrac{\phi}{s}\bm s\right] \cdot \left[\left(1 - \frac{h\kappa}{2}\right) \bm n - \frac{h_s}{2}\bm s\right] \\
&= \frac{1}{|\bm n_{\mathrm{U}}|}\left[\left(1-\frac{h\kappa}{2}\right)\pfrac{\phi}{n} - \frac{h_s}{2(1-h\kappa/2)} \pfrac{\phi}{s}\right].
\end{align*}
The component of the upper interface velocity in the centreline-normal direction, $v_{\mathrm{nU}}$, is then
\begin{subequations}
\label{eqs:interfaceVs}
\begin{equation}
v_{\mathrm{nU}} = \frac{|\bm n_{\mathrm{U}}| v_\mathrm{n}}{\bm n_{\mathrm{U}} \cdot \bm n} = \pfrac{\phi}{n} - \frac{h_s}{2(1-h\kappa/2)^2} \pfrac{\phi}{s}, \qquad n = \frac{h}{2}.
\end{equation}
Similarly, on the lower interface, the velocity in the centreline-normal direction is
\begin{equation}
v_{\mathrm{nL}} = \pfrac{\phi}{n} + \frac{h_s}{2(1+h\kappa/2)^2} \pfrac{\phi}{s}, \qquad n = -\frac{h}{2}.
\end{equation}
\end{subequations}

From the interface velocities \eqref{eqs:interfaceVs} we compute the centreline velocity and thickness evolution.  Let $\eta$ be a Lagrangian coordinate (such that $\eta$ is constant at a point on the curve $\bm x$ moving in the direction normal to the centreline); then
\[
\frac{D\bm x_{\mathrm{U}}}{Dt} = \frac{D\bar{\bm x}}{Dt} + \frac{h}{2} \frac{D\bm n}{Dt} + \frac{1}{2}\frac{Dh}{Dt}\bm n.
\]
Here the notation $D/Dt$ represents the Lagrangian derivative (holding $\eta$ constant).  Taking the centreline-normal component of each term in this equation, and noting that $D\bm n/Dt$ is orthogonal to $\bm n$, we find:
\[
v_{\mathrm{nU}} = v_\mathrm{n} + \frac{1}{2}\frac{Dh}{Dt}, \qquad v_{\mathrm{nL}} = v_\mathrm{n} - \frac{1}{2}\frac{Dh}{Dt},
\]
with the result for the lower interface derived in very similar fashion as that for the upper.  Arranging for $v_\mathrm{n}$ and $Dh/Dt$, and substituting \eqref{eqs:interfaceVs}, we then find expressions for the normal velocity and thickness evolution in the centreline-normal direction:
\begin{equation}
v_\mathrm{n} = 
\left\langle \pfrac{\phi}{n} \right\rangle - \frac{h_s}{4}\left[\frac{1}{(1-n\kappa)^2}\pfrac{\phi}{s}\right]_{-h/2}^{h/2}
, \qquad 
\frac{Dh}{Dt} = \left[\pfrac{\phi}{n}\right]_{h/2}^{h/2} - h_s\left\langle \frac{1}{(1-n\kappa)^2}\pfrac{\phi}{s}\right\rangle
\label{eq:generalFilamentEqs}
\end{equation}
where the notation $\langle f \rangle = (f|_{n=-h/2} + f|_{n=h/2})/2$ represents a value averaged between the two interfaces.

As is standard in free-boundary lubrication problems, it is useful to be able to express the thickness evolution in a conservative form.  To do so, we first note that Laplace's equation \eqref{eq:Laplace}, under the coordinate transformation \eqref{eq:intrinsicCoordinates}, becomes
\begin{equation}
\frac{1}{(1-n\kappa)^2}\frac{\partial^2\phi}{\partial s^2} + \frac{\partial^2\phi}{\partial n^2}+ \frac{n\kappa_s}{(1-n\kappa)^3}\frac{\partial\phi}{\partial s} - \frac{\kappa}{1-n\kappa}\pfrac{\phi}{n} = 0
\label{eq:intrinsicLaplace}
\end{equation}
which is equivalent to
\[
\pfrac{}{n}\left[(1-n\kappa)\pfrac{\phi}{n}\right] + \pfrac{}{s}\left[\frac{1}{1-n\kappa}\pfrac{\phi}{s}\right] = 0.
\]
Integrating over $n$ gives
\[
\left[\pfrac{\phi}{n}\right]_{-h/2}^{h/2} - h\kappa\left\langle \pfrac{\phi}{n} \right\rangle = -\int_{-h/2}^{h/2} \pfrac{}{s}\left[ \frac{1}{1-n\kappa}\pfrac{\phi}{s}\right] \, \mathrm dn
\]
so that, using \eqref{eq:generalFilamentEqs},
\begin{align*}
\frac{Dh}{Dt} - \kappa h v_\mathrm{n} &= -\int_{-h/2}^{h/2} \pfrac{}{s}\left[\frac{\phi_s}{1-n\kappa}\right]\,\mathrm dn - h_s\left\langle\frac{\phi_s}{(1-n\kappa)^2}\right\rangle + \frac{hh_s\kappa}{4}\left[\frac{\phi_s}{(1-n\kappa)^2}\right]_{-h/2}^{h/2} \\
&= -\int_{-h/2}^{h/2} \pfrac{}{s}\left[\frac{\phi_s}{1-n\kappa}\right]\,\mathrm dn - \frac{h_s}{2}\left.\frac{\phi_s}{1-n\kappa}\right|_{n=h/2} - \frac{h_s}{2} \left.\frac{\phi_s}{1-n\kappa}\right|_{n=-h/2}.
\end{align*}
Thus
\begin{equation}
\frac{Dh}{Dt} = \kappa h v_\mathrm{n} -\pfrac{}{s}\int_{-h/2}^{h/2} \frac{1}{1-n\kappa}\pfrac{\phi}{s} \,\mathrm dn.
\label{eq:hConserved}
\end{equation}
The first term on the right hand side of \eqref{eq:hConserved} represents the effect of dilation or compression of the film as its length changes, while the second term represents the contribution of flux along the filament.

The expressions \eqref{eq:generalFilamentEqs} and \eqref{eq:hConserved} are exact, but we have not yet determined the potential $\phi$.  We now formally take the lubrication approximation by substituting $n = \epsilon N$, $h = \epsilon H$, and $t = \epsilon T$, taking $\epsilon \ll 1$.  In this expansion we assume the curvature $\kappa = O(1)$ (or smaller), and the surface tension $\sigma = O(1)$.  Unlike most standard lubrication models where the leading order behaviour in $\epsilon$ is all that is required, in our case we will see that terms at order $\epsilon^2$ are necessary to fully regularise the instability, and so we will expand to this order.

Writing $\phi = \phi_0 + \epsilon \phi_1 + \epsilon^2\phi_2 + \ldots$ and substituting into Laplace's equation \eqref{eq:intrinsicLaplace}, we have
\begin{equation*}
\pfrac{^2\phi_0}{N^2} = 0 , \quad \pfrac{^2\phi_1}{N^2} = \kappa\pfrac{\phi_0}{N}, \quad \pfrac{^2\phi_2}{N^2} = \kappa\pfrac{\phi_1}{N} - \kappa^2N\pfrac{\phi_0}{N} - \pfrac{^2\phi_0}{s^2},
\end{equation*}
with pressure conditions from \eqref{eq:pressureBCL}, \eqref{eq:pressureBCU}:
\begin{align*}
\phi_0 &= \sigma\kappa, \quad \phi_1 = \sigma K_1, \quad \phi_2 = \sigma K_2, \qquad N= \frac{H}{2} \\
\phi_0 &= - P - \sigma\kappa, \quad \phi_1 = \sigma K_1, \quad \phi_2 = -\sigma K_2, \qquad N= -\frac{H}{2}.
\end{align*}
Here $\kappa$ is again the centreline curvature, and $K_1$ and $K_2$ are order $\epsilon$ and $\epsilon^2$ corrections to the curvature on the upper interface, respectively:
\begin{equation}
K_1 = \frac{H\kappa^2}{2} + \frac{H_{ss}}{2}, \qquad K_2 = \frac{H^2\kappa^3}{4} + \frac{HH_s\kappa_s}{4} + \frac{HH_{ss}\kappa}{2} + \frac{H_s^2\kappa}{8}
\label{eq:kCorrections}
\end{equation}
(the corrections on the lower interface are the same, with appropriate changes in sign).  Solving for each component of the potential $\phi$ term by term gives
\[
\phi_0 = V_0N - \frac{P}{2}, \quad \phi_1 = \frac{\kappa V_0}{2}\left(N^2 - \frac{H^2}{4}\right) + \sigma K_1, \quad \phi_2 = \frac{2\kappa^2V_0 - V_{0ss}}{6}\left(N^3 - \frac{H^2}{4}N\right) + \frac{2\sigma N}{H}K_2,
\]
where
\begin{subequations}
\label{eqs:filamentEqsEpsilon}
\begin{equation}
V_0 = \frac{P + 2\sigma \kappa}{H}
\end{equation}
is the leading-order centreline velocity.  Substitution of these expressions into the expansions of \eqref{eq:generalFilamentEqs} and \eqref{eq:hConserved} results in expressions for the rescaled normal velocity $V_\mathrm{n} = \epsilon^{-1}v_\mathrm{n}$ and $DH/DT = Dh/Dt$:
\begin{equation}
V_\mathrm{n} = V_0 + \epsilon^2\left[\frac{2H^2\kappa^2V_0 - H^2V_{0ss}}{12} - \frac{HH_sV_{0s}}{4} + \frac{2\sigma K_2}{H} \right]
\label{eq:VN}
\end{equation}
and
\begin{align}
\frac{DH}{DT} 
= \kappa H V_\mathrm{n} - \epsilon^2\pfrac{}{s}\left[-\frac{H^3\kappa_sV_0}{12} - \frac{H^2 H_s\kappa V_0}{4} + H\sigma K_{1s} \right] 
\label{eq:HT}
\end{align}
\end{subequations}

In \eqref{eq:VN}, \eqref{eq:HT}, the order of terms in the lubrication parameter $\epsilon$ is explicit.  We can of course formally remove the scaling by returning to $h, t$ variables, resulting in
\begin{subequations}
\label{eqs:filamentEqs}
\begin{align}
v_\mathrm{n} &= v_0 + \left(\frac{2h^2\kappa^2v_0 - h^2v_{0ss}}{12} - \frac{hh_s v_{0s}}{4}  + \frac{2\sigma \kappa_2}{h} \right) \label{eq:vn} \\
\frac{Dh}{Dt} &= \kappa h v_\mathrm{n} - \pfrac{}{s}\left[-\frac{h^3\kappa_s v_0}{12} - \frac{h^2h_s\kappa v_0}{4} + \sigma h \kappa_{1s}\right] 
\label{eq:DhDt}
\end{align}
with $v_0 = (P + 2\sigma\kappa)/h$ the leading-order velocity, and $\kappa_1 = \epsilon^{-1} K_1$, $\kappa_2 = \epsilon^{-2}K_2$ the rescaled versions of \eqref{eq:kCorrections}:
\begin{equation}
\kappa_1 = \frac{h\kappa^2}{2} + \frac{h_{ss}}{2}, \qquad \kappa_2 = \frac{h^2\kappa^3}{4} + \frac{hh_s\kappa_s}{4} + \frac{hh_{ss}\kappa}{2} + \frac{h_s^2\kappa}{8}.
\label{eq:kCorrectionsUnscaled}
\end{equation}
\end{subequations}

The higher order terms have been included in the above derivation as they are needed to fully regularise the instability.  This will become apparent when we examine the behaviour of the leading order system in Section \ref{sec:leadingOrder}.  In particular, the higher spatial derivative of thickness $h$ resulting from the interface curvature correction term $\kappa_{1}$ plays a crucial role. In the case where driving pressure $P$ and centreline curvature $\kappa$ both vanish, \eqref{eqs:filamentEqs} reduces to the well-known thin film equation
\begin{equation*}
\frac{\partial h}{\partial t} = -\frac{\sigma}{2}\pfrac{}{s}\left[h\pfrac{^3h}{s^3}\right],
\end{equation*}
where $h = h(s,t)$, and arclength $s$ becomes time-independent. This equation has been studied extensively in the context of droplet breakup in Hele--Shaw flow \citep{Almgren1996, Almgren1996b,Constantin1993,Dupont1993,Goldstein1993,Goldstein1995,Goldstein1998}.  In our case, this fourth order spatial derivative term stabilises high wavenumber perturbations, as is shown in the following stability analysis of the filament model.   

\subsection{Stability}

As with the full two-interface model \eqref{eqs:fullproblem}, the thin filament approximation \eqref{eqs:filamentEqs} has an exact solution comprising a straight filament of uniform thickness $h_0$ moving upward with speed $P/h_0$.  To test the stability of this straight filament, we write the centreline $\bar{\bm x}(s,t)$ in Cartesian coordinates as $y = f(x,t)$, write thickness $h$ as a function of $x$ and $t$, and perturb the straight filament:
\[
h(x,t) = h_0 + \tilde h\e^{\mathrm i k x + \lambda t}, \qquad f(x,t) = \frac{P}{h_0}t + \tilde f\e^{\mathrm i k x + \lambda t}.
\]
In the linear approximation, $D/Dt = \partial/\partial t + O(f_x^2)$, $s = x + O(f_x^2)$ and $\kappa = f_{xx} + O(f_x^2)$.  On substituting these expansions into \eqref{eqs:filamentEqs} we obtain the eigenvalue problem
\[
\lambda \begin{bmatrix} \tilde f \\ \tilde h \end{bmatrix} =
\begin{bmatrix}
-2\sigma k^2/h_0 & -P/h_0^2 + Pk^2/12 \\ -Pk^2 - Ph_0^2 k^4 /12  & -\sigma h_0 k^4/2
\end{bmatrix} \begin{bmatrix} \tilde f \\ \tilde h \end{bmatrix}.
\]
The eigenvalues are thus
\begin{equation}
\label{eq:filament_eigs}
\lambda = -\left(\frac{\sigma k^2}{h_0} + \frac{\sigma h_0 k^4}{4}\right) \pm \sqrt{\left(\frac{\sigma k^2}{h_0} + \frac{\sigma h_0 k^4}{4}\right)^2 + \frac{P^2k^2}{h_0^2} - \sigma^2 k^6 - \frac{P^2h_0^2 k^6}{144} }.
\end{equation}

We note that the linear stability of the filament model \eqref{eqs:filamentEqs} does not agree exactly with the full model, even for zero surface tension.  When $\sigma =0$, the error in the filament model is due to the final term $P^2h_0^2 k^6/144$, which arises from the product of the two higher order correction terms in the off-diagonal terms in the determinant.  This term is (implicitly) of  order $\epsilon^4$, which is of higher order than the model is accurate to.  If further terms were taken in the lubrication expansion, they would contribute further order $\epsilon^4$ terms which would presumably cancel with the term present here.

For nonzero surface tension $\sigma$, (\ref{eq:filament_eigs}) has the same limiting behaviours (\ref{eq:approx1}) and (\ref{eq:approx2}) as the full model, but differs slightly with (\ref{eq:approx3}).  In the latter case, both (\ref{eq:full_eigs}) and (\ref{eq:filament_eigs}) give $\lambda=O(1/h_0^3)$ as $h_0\rightarrow 0$ for $k=O(1/h_0)$, but with a different prefactor.  Thus, our thin filament approximation recovers the same leading-order linear stability behaviour as the full model for small and moderate wave numbers, which is all we can expect from such a lubrication model.  For very large wave numbers (i.e., very small wavelengths of perturbation), with $k\gg  1/h_0$, while the scalings between the eigenvalues of the full and filament model may differ, these modes of perturbation decay very quickly and therefore any differences in the models are of no practical consequence. 

In Fig.~\ref{fig:stability} we compare the eigenvalues of the thin filament approximation (\ref{eq:filament_eigs}) against those of the full problem (\ref{eq:full_eigs}).  For even moderately small thickness ($h_0 = 0.2$) and small surface tension ($\sigma = 0.1$), the agreement is excellent in the region in which eigenvalues are positive; the difference for negative values occurs for much larger $k$ than depicted.  These linear stability results do indicate, however, that the approximation will not be valid if surface tension is zero (or much smaller than the filament thickness), which is unsurprising given the implicit assumption that $\sigma = O(1)$ in the lubrication parameter $\epsilon$.

\section{Properties of the leading-order model}
\label{sec:leadingOrder}

In this section we consider the properties of the leading-order version of \eqref{eqs:filamentEqs}, that is, neglecting terms of order $\epsilon^2$:
\begin{equation}
v_\mathrm{n} = v_0 = \frac{P + 2\sigma \kappa}{h}, \qquad \frac{Dh}{Dt} = \kappa h v_0,
\label{eq:leadingOrderIntrinsic}
\end{equation}
where $\kappa$ is the centreline curvature.  In the leading-order model \eqref{eq:leadingOrderIntrinsic}, there is no fluid flow tangent to the centreline; the change in filament thickness is due purely to the stretching or compression of the filament.  

Potential issues with the leading-order model can first be observed in the linear stability analysis of a flat filament.  The stability analysis of the leading-order model \eqref{eq:leadingOrderIntrinsic} is the same as for the higher order model with the higher-order terms absent; the eigenvalues of this leading-order system are thus
\begin{equation}
\label{eq:notransverse_eigs}
\lambda = -\frac{\sigma k^2}{h_0} \pm \sqrt{\left(\frac{\sigma k^2}{h_0}\right)^2 + \frac{P^2k^2}{h_0^2}}.
\end{equation}
For small $k$, (\ref{eq:notransverse_eigs}) coincides with the leading-order behaviour (\ref{eq:approx1}), while for $k=O(1)$, (\ref{eq:notransverse_eigs}) is no longer a reasonable approximation for the full model.  Indeed, unlike in \eqref{eq:filament_eigs}, where both eigenvalues becomes negative for sufficiently large wavenumber $k$, in \eqref{eq:notransverse_eigs}, one eigenvalue tends to a positive, constant value as $k\to\infty$:
\[
\lambda \sim \frac{P^2}{2\sigma h_0}, \qquad k \to \infty
\]
(see Fig.~\ref{fig:stability}).  Thus the system is (significantly) unstable to perturbations at arbitrarily small spatial scales, even in the presence of surface tension.  While this analysis does not suggest the leading-order problem is technically ill-posed if $\sigma > 0$ (the eigenvalues do not become arbitrarily large), this large-$k$ behaviour strongly suggests that singularities in curvature will generically form \citep[see for example][]{Dallaston2014}.  We will confirm the self-similar formation of curvature singularities in Section \ref{sec:leadingOrderSingularityFormation}.

\subsection{The unregularised (zero surface tension) leading-order model}
\label{sec:unregularized}

In the case $\sigma = 0$, the leading order model \eqref{eq:leadingOrderIntrinsic} reduces to that considered by \citet{Farmer2006}.  In this case the problem is indeed ill-posed; the eigenvalues \eqref{eq:notransverse_eigs} are $\lambda = \pm (P/h_0)k$, as are the eigenvalues of the full two-interface problem \eqref{eq:full_eigs}, so one eigenvalue is arbitrarily large as $k\to\infty$.  Solutions that exhibit this ill-posedness may be constructed by showing that the centreline is given by level curves of a harmonic function.  We summarise the approach of \citet{Farmer2006} and further examine this approach here.

Assuming the filament centreline $(x(\eta,t),y(\eta,t))$ and the thickness $h(\eta,t)$ are parametrised by a Lagrangian coordinate $\eta$, then
\[
\frac{\partial x}{\partial t}=-\frac{P y_\eta}{h\sqrt{x_\eta^2 + y_\eta^2}},
\qquad
\frac{\partial y}{\partial t}=\frac{P x_\eta}{h\sqrt{x_\eta^2 + y_\eta^2}}.
\]
For $\sigma=0$, the evolution of $h$ \eqref{eq:leadingOrderIntrinsic} is equivalent to conservation of mass between a point $\eta$ and reference point $\eta_0$.  That is, we may define an area function $A(\eta)$, such that
\begin{equation}
A(\eta)=\int_{\eta_0}^\eta h(\bar{\eta},0)\,\mathrm{d}\bar{\eta}
=\int_{\eta_0}^\eta h(\bar{\eta},t)\sqrt{x_\eta^2 + y_\eta^2}
\,\mathrm{d}\bar{\eta}
\label{eq:areafn}
\end{equation}
is constant in time on a point on the centreline moving with its normal velocity.  Choosing to scale time such that $P=1$, we arrive at the following:
\begin{equation}
\label{eq:CR}
\frac{\partial x}{\partial t}=-\frac{\partial y}{\partial A},
\qquad
\frac{\partial y}{\partial t}=\frac{\partial x}{\partial A}.
\end{equation}
These are Cauchy--Riemann equations relating $(x,y)$ to $(A,t)$.  This line of argument was used by \citet{Farmer2006} to demonstrate that $w = A + \mathrm i t$ must be an analytic function of the complex spatial variable $z=x+\mathrm{i}y$; thus, for a given time $t$, the centreline is the level curve of the harmonic function $t(x,y)$.  

Given the definition of $A$ \eqref{eq:areafn}, the thickness $h$ may be calculated by:
\begin{equation}
\label{eq:exacth}
h = \frac{A_xx_\eta + A_y y_\eta}{\sqrt{x_\eta^2 + y_\eta^2}} = \frac{A_x t_y - A_y t_x}{\sqrt{t_x^2 + t_y^2}} = \sqrt{t_x^2 +t_y^2} = \left|w'(z)\right|.
\end{equation}
This thickness will go to zero at a critical point $z_c$ where $w'(z_c) = 0$.  As this is also a point where the conformal map between $w$ and $z$ ceases to be smooth, we expect to see a singularity in the curvature in the centreline there.  The preimage of the straight line $w = A + \ii t_c$ that passes through the critical point is the centreline in the $z$-plane; assuming $w''(z_c)\neq 0$ the centreline must therefore have a corner with an angle of $\pi/2$ at $z_c$ (and not a cusp, as suggested by \citet{Farmer2006}).  If the initial condition is such that $w''(z_c)$ also vanishes but the third derivative is nonzero, the corner angle is $\pi/3$, and so on.  

\begin{figure}
\centering
\includegraphics{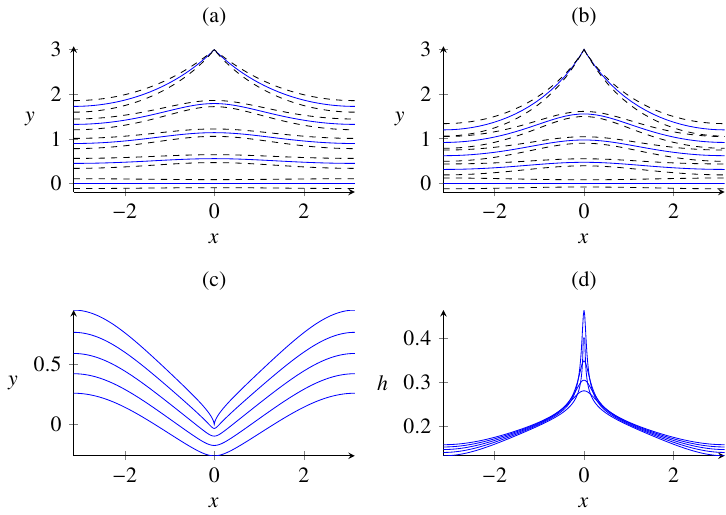}
\caption{
Exact solutions to the thin filament equation in the absence of surface tension. (a) the solution \eqref{eq:exactsol1} that evolves from an initial condition that results in corner formation with the generic angle of $\pi/2$, and (b) the solution \eqref{eq:exactsol2} that evolves from an initial condition that results in corner formation with a non-generic angle of $\pi/3$.  The parameter values are $a=0.1$, $\delta=0.2$.  In both of these examples, the centrelines are plotted in solid blue while the black dashed lines are the upper and lower interfaces (found by adding or subtracting half the thickness $h$ in the normal direction).  (c), the profiles, and (d) the thickness, for the example \eqref{eq:exactsol3} with $\delta=0.2$ and $a=0.5$, which forms a zero angle cusp with infinite thickness.}
\label{fig:exact_solutions}
\end{figure}

As an example, consider an initial condition with the centreline on $y=0$ and initial thickness given by $h(x,0) = \delta[1-a\cos x]$, with $\delta > 0$ and $0 < a < 1$.  This initial condition corresponds to an initially horizontal filament that is thinner near $x=0$.  We thus have $A = \delta[x - a\sin x]$ at $t=0$ (determined by choosing our reference point $\eta_0$ to lie on $x=0$), and, analytically continuing into the complex plane:
\begin{equation}
A + \ii t = \delta(z - a\sin z).
\end{equation}
Taking the imaginary part, we find that the centreline location is given implicitly by
\begin{equation}
\label{eq:exactsol1}
t = \delta(y - a\cos x\sinh y).
\end{equation}
The critical point occurs for $z_c = \ii\cosh^{-1}(1/a)$ and time $t_c = \delta(\cosh^{-1}(1/a) - \sqrt{1-a^2})$.  The centreline profiles of this solution, along with the upper and lower interfaces (found by adding and subtracting half the thickness $h$ \eqref{eq:exacth}, respectively, in the normal direction) are plotted in Fig.~\ref{fig:exact_solutions}a, showing the formation of the $\pi/2$ angle as $t\to t_c^-$. 

 As an example of an initial condition that leads to a non-$\pi/2$ angle, consider
 \[
 h(x,0) = \delta[1 - a\cos x]^2
 \]
which on integrating results in
\begin{equation}
A + \ii t = \delta \left[\left(1 + \frac{a^2}{2}\right)z + \frac{a^2}{4}\sin 2z - 2a\sin z\right].
\label{eq:exactsol2}
\end{equation}
Again, the centreline is determined by taking the contours of the imaginary part of this function.  In this case, since $w'(z_c) = w''(z_c) = 0$, the corner angle at the critical point where $h\to 0$ is $\pi/3$.  Profiles of this solution are shown in Fig.~\ref{fig:exact_solutions}b.

The previous two exact solutions show only a subset of the possible singular behaviours of the ill-posed zero-surface tension model; as centrelines are given by level curves of a harmonic function $t = \Im(w(z))$, with thickness given by \eqref{eq:exacth}, a variety of other solution behaviour is possible.  One such possibility is a zero-angle cusp that results from a square root singularity, an example of which is the complex function 
\begin{equation}
A + \ii t = w(z) = \delta \left(z - a\ii\sqrt{1-\e^{-\ii z}}\right).
\label{eq:exactsol3}
\end{equation}
This function has a square root singularity at $z=0$, as well as the appropriate periodicity.  The level curve that passes through this singularity occurs at $t=0$, so as $t\to 0^{-}$, a zero angle, backward-facing cusp forms on the interface.  The thickness $h$, which is given by $|w'(z)|$, becomes infinite at the cusp, so that the interface at the cusp becomes stationary.  The profiles and thickness corresponding to \eqref{eq:exactsol3} are shown in Fig.~\ref{fig:exact_solutions}c--d, respectively, for the parameter values of $\delta = 0.2$ and $a=0.5$.

\subsection{Singularity formation in the leading-order model with surface tension}
\label{sec:leadingOrderSingularityFormation}

We now further consider the leading order-model \eqref{eq:leadingOrderIntrinsic} in the presence of surface tension.  As eigenvalues \eqref{eq:notransverse_eigs} are positive for arbitrarily large wavenumber, we expect the possibility of singularities in curvature, and this does indeed generically occur, as we demonstrate through both the construction of similarity solutions, and corroborate with numerical simulations.

To establish the existence of curvature singularities we perform a self-similar analysis.  This analysis is simplest to carry out in Cartesian coordinates. Let the thickness $h = h(x,t)$ be a function of $x$ and $t$, and define the centreline to be given by the function $y = f(x,t)$; we then have
\[
v_0 = \frac{1}{\sqrt{1+f_x^2}} \pfrac{f}{t}, \qquad \frac{Dh}{Dt} = \pfrac{h}{t} - \frac{v_0 f_x}{\sqrt{1+f_x^2}} \pfrac{h}{x}
\]
where $\partial/\partial t$ is the time derivative with $x$ held constant.  Using \eqref{eq:leadingOrderIntrinsic} then results in
\[
\pfrac{f}{t} = \frac{\sqrt{1+f_x^2}}{h}\left(P + 2\sigma \frac{f_{xx}}{(1+f_x^2)^{3/2}}\right), \quad \pfrac{h}{t} = \left(\frac{f_x h_x}{1+f_x^2} + \frac{f_{xx}h}{(1+f_x^2)^2}\right) \pfrac{f}{t}.
\]
Further simplification is achieved by defining the filament thickness in the $y$ direction $\hy(x,t) = h\sqrt{1+f_x^2}$ (as opposed to the thickness $h$ in the centreline-normal direction), which results in the system of equations
\begin{equation}
\pfrac{f}{t} = \frac{1+f_x^2}{\hy}\left(P + 2\sigma \frac{f_{xx}}{(1+f_x^2)^{3/2}}\right), \qquad \pfrac{\hy}{t} = \pfrac{}{x}\left[f_x\left(P + 2\sigma\frac{f_{xx}}{(1+f_x^2)^{3/2}} \right) \right].
\label{eq:leadingOrderCartesian}
\end{equation}
By introducing $\hy$, \eqref{eq:leadingOrderCartesian} has been written in conservative form, which highlights the fact that mass is indeed conserved in this system.

Assume a singularity occurs at a time $t=t_c$ at $x=x_c$, and let:
\begin{equation}
f \sim f_0(t) + (t_c-t)^\alpha F(\xi), \qquad \hy \sim (t_c-t)^\beta H(\xi), \qquad \xi = \frac{x-x_c}{(t_c-t)^\gamma},
\end{equation}
where the similarity exponents $\alpha,\beta,\gamma$ are to be determined.  Furthermore, we assume the profile and thickness are symmetric in $x$.  Assuming that $\alpha > 1$, the dominant term in the velocity $f_t$ is $\dot f_0 = \dot f_0(t_c)$.  Thus, on balancing terms we find $\beta = -1$ and $\alpha = 2\gamma -1$, with $\gamma$ being undetermined (a second-kind self-similarity). Given $\alpha > \gamma$ (which we will check for consistency after the fact), the dominant terms in \eqref{eq:leadingOrderCartesian} become
\[
\dot f_0 = 2\sigma \frac{F''}{H}, \qquad H + \gamma\xi H' = 2\sigma[F'F'']'.
\]
These equations can be further scaled to remove $\dot f_0$ and $\sigma$.  Let $F = (\dot f_0)^{-1}\hat F$ and $H = 2\sigma(\dot f_0)^{-2} H$, then
\[
\hat H = \hat F'', \qquad \hat H + \gamma\eta \hat H' = [\hat F'\hat F'']'.
\]
Let $u = \hat F'$ and eliminate $\hat F$, so that $u$ satisfies the second-order equation
\[
u'' = \frac{u'(u'-1)}{\gamma\xi - u}.
\]
For symmetry we require $u$ to be odd in $\xi$; thus, when $\xi=0$, $u=0$, which is therefore a singular point of the ODE.  By expanding near this point we find that the similarity exponent $\gamma$ is uniquely specified by the first odd power in the expansion greater than unity:
\[
u \sim \xi - C\xi^n, \qquad \xi \to 0, \qquad n = 3, 5, \ldots
\]
where $C$ is an arbitrary scaling factor.  Expanding near the singular point results in $\gamma = n/(n-1) > 1$, which is consistent with the assumptions made on the similarity exponents above.

Given $\gamma = n/(n-1)$, the equation for $u$ can be solved implicitly by letting $u$ be the independent variable, which allows us to construct parametric solutions for $\hat F$ and $\hat G$:
\begin{equation}
\label{eq:blowup_sols}
\xi = u + Cu^n, \qquad \hat H = \frac{1}{1+nCu^{n-1}}, \qquad \hat F = \frac{u^2}{2} + \frac{n}{n+1} C u^{n+1}.
\end{equation}
While $n$ can be any odd number $\geq 3$, dependent on the initial condition, the most generic case will be $n=3$, in which case $\gamma = 3/2$ and $\alpha=2$.  The constant $C$ is arbitrary, due to scale invariance in the equations \eqref{eq:leadingOrderCartesian} in the limit that curvature dominates over the driving pressure $P$.  In a given case, the scale $C$, velocity $\dot f_0$, and exponent $n$ will all depend on the initial condition of the time-dependent problem.

We provide numerical evidence for the curvature singularity formation by numerically solving the system \eqref{eq:leadingOrderCartesian}.  This computation is performed in MATLAB using finite difference discretisation along with MATLAB's \texttt{ode15s} algorithm for time-stepping.  Parameters $P = 1$, $\sigma = 0.5$ and an initial condition of $f(x,0) = -\cos(x)$ and $h(x,0) = 1$ is chosen in order to start with high curvature near $x=0$, where the singularity will occur.

\begin{figure}
\centering
\includegraphics{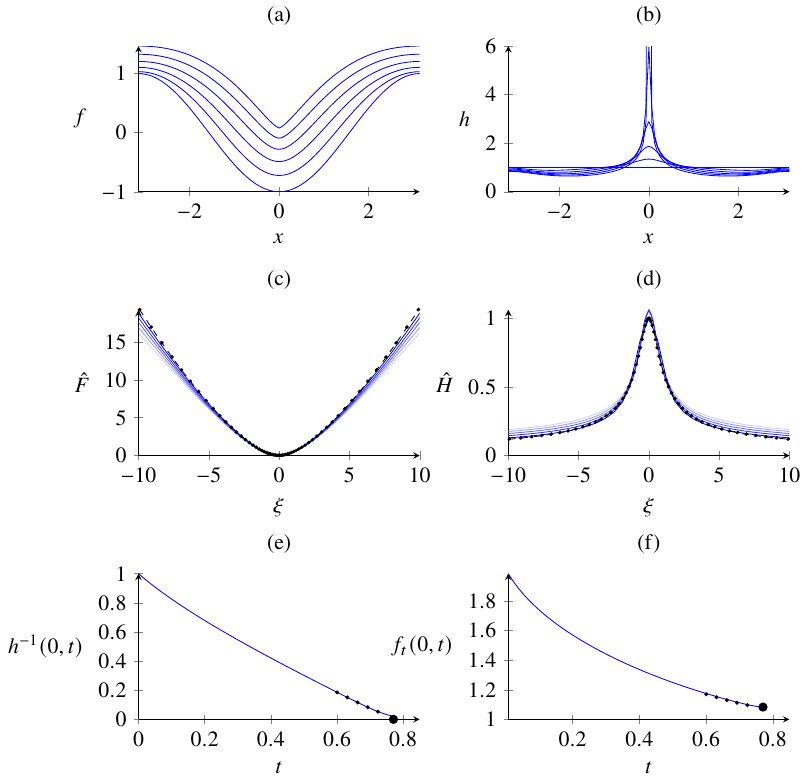}
\caption{
Curvature singularity formation in the leading-order Hele--Shaw filament model \eqref{eq:leadingOrderCartesian} for initial condition $f(x,0) = -\cos(x)$, and $\hy(x,0) = 1$.  (a) centreline profiles $f(x,t)$ approaching a curvature singularity and (b) filament thickness becoming unbounded at singularity time of $t_c \approx 0.77$.  These profiles (solid lines) collapse onto similarity profiles (c) $\hat F(\xi)$ and (d) $\hat H(\xi)$, which asymptotically match the exact similarity solution \eqref{eq:blowup_sols} (dotted lines) as $t\to t_c$.  The scaling factor $C\approx 0.3$ is fit to the profiles.  (e) The singularity time (large dot) is found by fitting a straight line approximation to the reciprocal of the maximum thickness near the singularity time, while (f) the speed of the centreline $\dot f_0 = f_t(0, t_c)$ (large dot) at the singularity is similarly estimated.  The five smaller points in (e,f) are the times at which the scaled profiles are plotted in (c,d).}
\label{fig:curvature_singularity}
\end{figure}

In Fig.~\ref{fig:curvature_singularity} we plot the results of this numerical computation.  The singularity time $t_c$ is estimated by fitting a straight line through $h_\mathrm{max}(t)^{-1}=(\max_x h(x,t))^{-1}$, which occurs at $x=0$, and should (according to the analysis above) go to zero linearly.  The centreline velocity $\dot f_0(t)$ at the maximum thickness is observed to tend to a nonzero constant, from which $\dot f_0$ is estimated.
Scaling the profiles near the singularity time into similarity variables $\xi, \hat F, \hat H$, we observe collapse.  The exact similarity profiles \eqref{eq:blowup_sols}, with a suitable fitted value of the arbitrary constant $C$, match well with the numerical profiles.

The curvature singularity exhibited by this model is weaker than a corner singularity, in that as $\alpha > \gamma$, the first derivative goes to zero in the neighbourhood of the singularity, even while the curvature becomes unbounded.  It is also interesting to note that the singularity is not dependent on (and does not require) the driving pressure $P$; its cause is the presence of surface tension pulling regions of high positive curvature inward, concentrating the thickness at a single point. It is thus only present as the leading-order model \eqref{eq:leadingOrderIntrinsic} has a surface tension-dependent velocity, but no regularising term that penalise higher derivatives of thickness, as appears in the second-order model in \eqref{eq:DhDt}.

\subsection{Quasi-travelling wave solutions}
\label{sec:travwave}

The system \eqref{eq:leadingOrderCartesian}, while insufficiently regularised to simulate the full dynamics of a thin filament, does exhibit a quasi-travelling wave solution of the form (in Cartesian variables) 
\begin{equation}
f = -B\log(t_0-t) + F(x), \qquad \hy(x,t) = (t_0-t)H(x), 
\label{eq:travWaveAnsatz}
\end{equation}
where $t_0$ is a finite time.  A solution of this form is similar to, but not exactly, a travelling wave, as the centreline has a fixed shape but moves to infinity (with speed unbounded) as $t\to t_0$, while the thickness linearly decreases to zero.  The parameter $B$ is the analogue of a travelling wave speed.  Solutions to \eqref{eq:leadingOrderCartesian} of this form generalise the `grim-reaper' solutions to the zero surface tension problem found in \citet{Farmer2006}, and are also candidates for asymptotically valid solutions to the higher-order system \eqref{eqs:filamentEqs}, since as the thickness goes to zero, we would expect the higher-order terms to vanish at faster rate than the leading-order terms.  We will see in Section \ref{sec:numerics} that these quasi-travelling waves do not appear to be attractors, but will compute them here for completeness.

\begin{figure}
\centering
\includegraphics{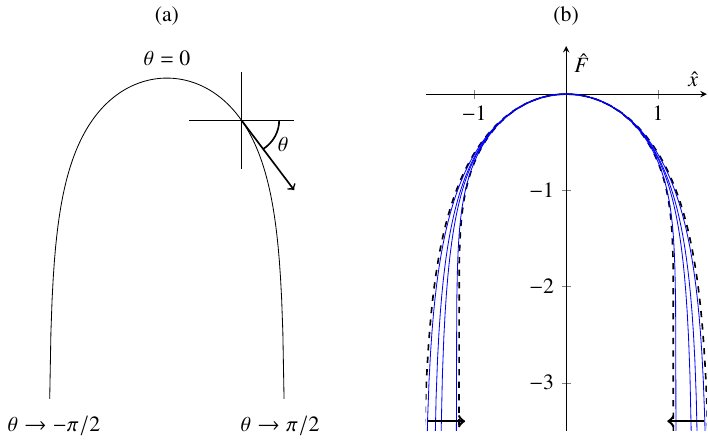}
\caption{(a) Depiction of the curvature-angle coordinate system used to solve the quasi-travelling wave solutions \eqref{eq:travwaveK}.  (b) Quasi-travelling wave solutions to \eqref{eq:travwaveK}, for $\epsilon=0$ and $\epsilon\to\infty$ (dashed curves), and $\hat\sigma \in \{0.1, 0.5, 1, 10\}$ (solid curves).  The arrows indicate direction of increasing $\hat\sigma$.}
\label{fig:travwave}
\end{figure}

On substitution of the ansatz \eqref{eq:travWaveAnsatz} into the leading-order model in Cartesian form \eqref{eq:leadingOrderCartesian}, and scaling the variables according to
\[
H = \frac{P}{B}\hat H, \qquad F = B\hat F, \qquad x = B\hat x
\]
we obtain
\begin{equation}
-\left[\hat F'\left(1 + \hat\sigma \frac{\hat F''}{(1+\hat F'^2)^{3/2}}\right)\right]' = (1+\hat F'^2)\left(1 + \hat\sigma \frac{\hat F''}{(1+\hat F'^2)^{3/2}}\right), \qquad \hat\sigma = \frac{2\sigma}{BP}.
\label{eq:travwaveF}
\end{equation}
This equation may be solved numerically directly, but to further simplify we cast it into the following curvature-angle formulation.  Let $\theta = -\tan^{-1} \hat F'$ be the angle between the tangent to the centreline and the $x$-axis (counting positive for negative $F'$), and scaled curvature $K = B^{-1}\kappa = \hat F''/(1+\hat F'^2)^{3/2}$ (see Fig.~\ref{fig:travwave}a).  We then obtain the first-order equation
\begin{equation}
\frac{\mathrm dK}{\mathrm d\theta} = -\frac{1 + \hat\sigma K}{\hat\sigma K \sin\theta}\left(1+\frac{K}{\cos\theta}\right).
\label{eq:travwaveK}
\end{equation}
For a semi-infinite curve, the appropriate interval is  $-\pi/2 < \theta < \pi/2$, where the nose at $\theta=0$ is a singular point, at which we require $K(0) = -1$.  In the limit of the tail of the travelling wave $\theta \to \pi/2$, we must have  $K \sim \theta-\pi/2 \to 0$.

We solve equation \eqref{eq:travwaveK} numerically for different values of $\hat\sigma$, and reconstruct the $\hat x, \hat F$ coordinates, using
\begin{equation}
\frac{\mathrm d \hat x}{\mathrm d\theta} = -\frac{\cos\theta}{K}, \qquad \frac{\mathrm d\hat F}{\mathrm d\theta} = \frac{\sin\theta}{K}.
\label{eq:constructXAndF}
\end{equation}
The $\hat\sigma=0$ and $\hat\sigma\to\infty$ limits are amenable to exact solutions.  If $\hat\sigma=0$ then $K = K_0 = -\cos(\theta)$, which is the `grim reaper' solution found by \citet{Farmer2006} (also relevant for the zero-surface-tension Saffman-Taylor finger \citep{Saffman1958} and for curve shortening flow \citep{Angenent1991}), equivalent to a centreline profile $\hat F = \ln(\cos\hat x)$.  On the other hand, as $\hat\sigma \to\infty$ the equation for $K$ becomes linear, and has exact solution
\begin{equation}
K = K_\infty = -\cot\theta\log\left(\frac{\cos(\theta/2) + \sin(\theta/2)}{\cos(\theta/2) - \sin(\theta/2)}\right).
\label{eq:Kinf}
\end{equation}

Plots of these quasi-travelling wave solutions are shown in Fig.~\ref{fig:travwave}b.  The effect of $\hat\sigma$, and thus of surface tension, is weak, as all solutions are bounded between the $\hat\sigma =0 $ and $\hat\sigma \to\infty$ limits.  The scaled curvature at the nose is required to be $-1$ in all cases, and thus returning to the unscaled system, we expect the curvature at the nose to be $-1/B$, that is, inversely proportional to the wave speed parameter.  

The corresponding scaled thickness in the $y$ direction is $\bar H = \sec^2\theta(1+\hat\sigma\kappa)$.  The time-dependent thickness in the centreline-normal direction is thus
\[
h = \frac{P(t_0-t)}{B\cos\theta} (1+\hat\sigma \kappa).
\]
At a fixed time $t_0$ then, the thickness is unbounded in the tails as $\theta\to \pm\pi/2$.  If, however, we consider a fixed value of $y$ (in the non-travelling coordinate system) as the singular time $t_0$ is approached, then from \eqref{eq:travWaveAnsatz} and \eqref{eq:constructXAndF}, the corresponding value of $\theta$ behaves as
\[
\pi/2 - \theta \sim (t_0-t)\e^{y/B}, \qquad t\to t_0.
\]
Thus $h = O(\e^{-y/B})$ is bounded for a fixed value of $y$ as $t\to t_0$.  The exponential decay in $y$ of the thickness also means that a quasi-travelling wave solution does not require infinite mass to form, even as it length is unbounded in the $y$-direction as $t\to t_0$.

In Section \ref{sec:numerics} we will simulate the second order thin filament model, including with initial conditions close to a quasi-travelling wave solution.  These simulations will demonstrate that these quasi-travelling wave solutions do not appear to be stable.

\section{Numerical simulation and comparison}
\label{sec:numerics}

In this section we describe numerical simulations of the second-order thin filament model \eqref{eqs:filamentEqs}, and compare against simulations of the full two-interface problem \eqref{eqs:fullproblem}.

Our solution method for the thin filament model \eqref{eqs:filamentEqs} is a front tracking method, whereby the centreline is represented by $N$ points $\bm x_j = (x_j, y_j), j=1,\ldots, N$.  The thickness also has a value $h_j$ at each point. At a given time, the normal vector, curvature, and arclength derivatives of $h$ and related quantities in  \eqref{eqs:filamentEqs} are calculated using a central finite difference scheme.  The points are then moved in time in the normal direction with velocity given by \eqref{eq:vn} (consistent with \eqref{eq:DhDt} being Lagrangian evolution equation for $h$) using MATLAB's \texttt{ode15s}.  Since moving points with normal velocity results in highly unevenly spaced points on the centreline, we remesh onto an evenly spaced grid when the ratio between minimum and maximum node spacing drops below a threshold value.  We typically use $2\,000$--$5\,000$ points (increasing this resolution does not have an appreciable effect in the simulations reported here) and remesh when the minimum-to-maximum node spacing is less than $0.8$. 

\subsection{Validation against solution of the two-interface problem}
\label{sec:levelset}

To validate solutions of the thin filament model \eqref{eqs:filamentEqs} it is valuable to compare it to solutions of the full two-interface system \eqref{eqs:fullproblem}, which is a more challenging numerical problem.  We solve \eqref{eqs:fullproblem} using the numerical framework proposed by \citet{Morrow2021,Morrow2023}, which we briefly summarise here. The framework is based on the level set method, in which we represent each interface, $f_\mathrm{L}$ and $f_\mathrm{U}$, as the zero level set of the associated level set functions $\psi_{\mathrm{L}}$ and $\psi_{\mathrm{U}}$. Each level set function is chosen such that the viscous fluid will occupy the region where both $\psi_{\mathrm{L}}$ and $\psi_{\mathrm{U}} > 0$; otherwise the region is filled with inviscid fluid. Both level set functions are updated according to
\begin{align}
	\frac{\partial \psi_{\mathrm{L}}}{\partial t} + f_\mathrm{L} | \nabla \psi_{\mathrm{L}}| = 0 \quad \textrm{and} \quad \frac{\partial \psi_{\mathrm{U}}}{\partial t} + f_\mathrm{U} | \nabla \psi_{\mathrm{U}}| = 0, \label{eq:LevelSetEqns}
\end{align}
where
\begin{align}
	f_\mathrm{L} = \nabla \phi \cdot \bm{n}_\mathrm{L} \quad \textrm{and} \quad f_\mathrm{U} = \nabla \phi \cdot \bm{n}_\mathrm{U}, \label{eq:SpeedFnc}
\end{align}
and $\bm{n}_\mathrm{L}  = \nabla \psi_{\mathrm{L}} / |\nabla \psi_{\mathrm{L}}|$ and $\bm{n}_\mathrm{U}  = \nabla \psi_{\mathrm{U}} / |\nabla \psi_{\mathrm{U}}|$ are the unit (outward) normals. We discretise the spatial derivatives in \eqref{eq:LevelSetEqns} using a second-order essentially non-oscillatory scheme and integrate in time using second-order total-variation-diminishing Runge--Kutta with time step $\Delta t = 10^{-5}$. We perform simulations on the computational domain $-\pi \le x \le \pi$ and $-0.5 \le y \le 4$, which is discretised into equally spaced nodes with mesh size $\Delta x = 2 \pi / 750$. Simulations are concluded when the minimum distance between the two interfaces is less than $3 \Delta x$. Further, we periodically perform reinitialisation to maintain both $\psi_{\mathrm{L}}$ and $\psi_{\mathrm{U}}$ as signed distance functions.  The details of this reinitialisation procedure are described in \citet{Morrow2021}.

We solve \eqref{eq:Laplace} for the velocity potential $\phi$ via a finite difference stencil. Following \citet{Chen1997}, we modify the stencil at nodes adjacent to either interface, corresponding to where $\psi_{\mathrm{L}}$ or $\psi_{\mathrm{U}}$ changes sign, by imposing a ghost node on the interface to incorporate the appropriate dynamic boundary conditions \eqref{eq:pressureBCL} and \eqref{eq:pressureBCU}. Here, $\kappa_\mathrm{L} = \nabla \cdot \bm{n}_\mathrm{L}$ and $\kappa_\mathrm{U} = \nabla \cdot \bm{n}_\mathrm{U}$. By solving for $\phi$, we can compute $f_\mathrm{L}$ and $f_\mathrm{U}$ from \eqref{eq:SpeedFnc}. These choices of $f_\mathrm{L}$ and $f_\mathrm{U}$ satisfy the kinematic boundary conditions \eqref{eq:kinematicBCs} and gives a continuous expression for $f_\mathrm{L}$ and $f_\mathrm{U}$ in the region occupied by the viscous fluid $\bm{x} \in \Omega$. However, to solve \eqref{eq:LevelSetEqns}, we require expressions for $f_\mathrm{L}$ and $f_\mathrm{U}$ over the entire computational domain. To extend our expressions for $f_\mathrm{L}$ and $f_\mathrm{U}$ into the gas regions, we follow \citet{Moroney2017} by solving the biharmonic equation
\begin{align}
	\nabla^4 f_\mathrm{L} = 0 \quad \textrm{and} \quad \nabla^4 f_\mathrm{U} = 0 \quad \textrm{in} \quad \bm{x} \in \mathbb{R}^2 \backslash \Omega.
\end{align}
By doing so, we obtain smooth continuous normal velocities over the entire computational domain, allowing us to solve \eqref{eq:LevelSetEqns} for $\psi_{\mathrm{L}}$ and $\psi_{\mathrm{U}}$.

\begin{figure}
\centering
\includegraphics{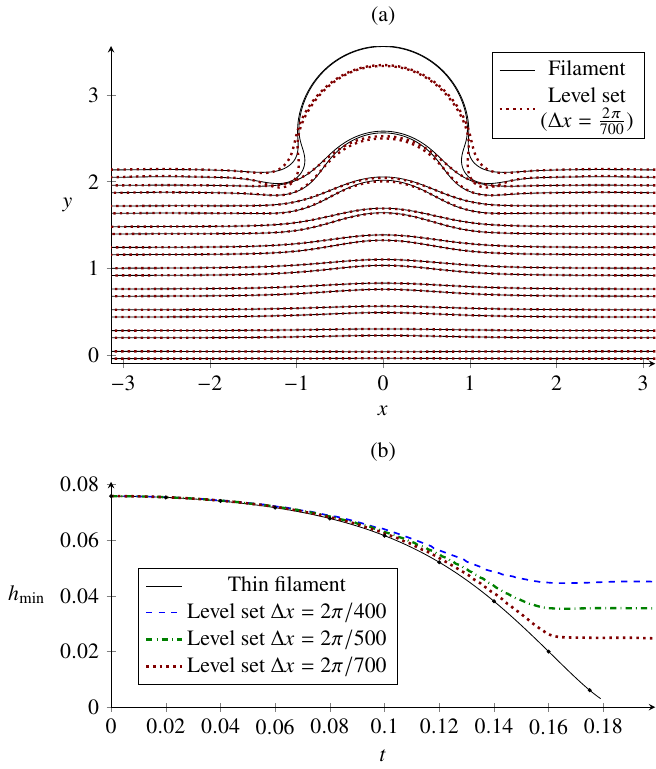}
\caption{(a) A comparison between the interfaces predicted from numerical solution of the filament model, and those from numerical solution of the full problem using the level set method described in Section \ref{sec:levelset}.   The initial condition is given by \eqref{eq:levelSetICs} with pressure and surface tension parameters $P=1$ and $\sigma = 0.1$, respectively. For clarity, only the interfaces of the filament model, found from the actual variables of centreline and thickness, are plotted. (b) A comparison of the minimum thickness $h_\mathrm{min}$ (which occurs at the nose $x=0$ between the filament model and level set simulations, as the resolution of the level set simulation is increased.  For smaller $\Delta x$, the level set method tends to the thin filament result, while the filament thickness in the level set method saturates at a value that depends on the numerical resolution.  The points marked on the filament curve correspond to the times at which the profiles are plotted in (a).}
\label{fig:levelSetComparison}
\end{figure}

In Fig.~ \ref{fig:levelSetComparison} we compare the results of the filament model and full two-interface model, with initial conditions:
\begin{equation}
	f_\mathrm{U}(x, 0) = 1/24 - 0.00375 \sech^2 x, \qquad
	f_\mathrm{L}(x, 0) = -1/24 + 0.00375 \sech^2 x.
	\label{eq:levelSetICs}
\end{equation}
These initial conditions correspond to an initial centreline location of $y=0$ and an initial thickness $h(x,0) = 1/12 - 0.0075\sech^2 x$, which is an almost flat filament with a small thinner region near the centre of the channel, $x=0$.  As demonstrated in Fig.~\ref{fig:levelSetComparison}a, the agreement between the two methods is initially very good, and only starts to disagree quantitatively when the filament thins near the central region, leading to a large increase in velocity.  This is to be expected, as the level set method will become inaccurate when the filament thickness becomes too thin to capture accurately using a level set function on discretised mesh.  To demonstrate this point, in Fig.~\ref{fig:levelSetComparison}b we plot the minimum thickness $h_\mathrm{min}(t)$ against time, for level set computations of increasing resolution.  We see that the level set result does approach that of the filament model as the resolution is increased (of course, we do not expect the results to be identical due to the approximation made in deriving the thin filament model \eqref{eqs:filamentEqs}, but this error is much smaller than the achievable discretisation error in the level set simulation).  Ultimately in the level set simulations, the thickness saturates at a value dependent on the resolution.  This limitation means the level set simulations ultimately underestimate the speed at which the filament advances when it becomes very thin, as can be observed in the profiles shown in Fig.~\ref{fig:levelSetComparison}a.

\subsection{Numerical results for late times}
\begin{figure}
\centering
\includegraphics{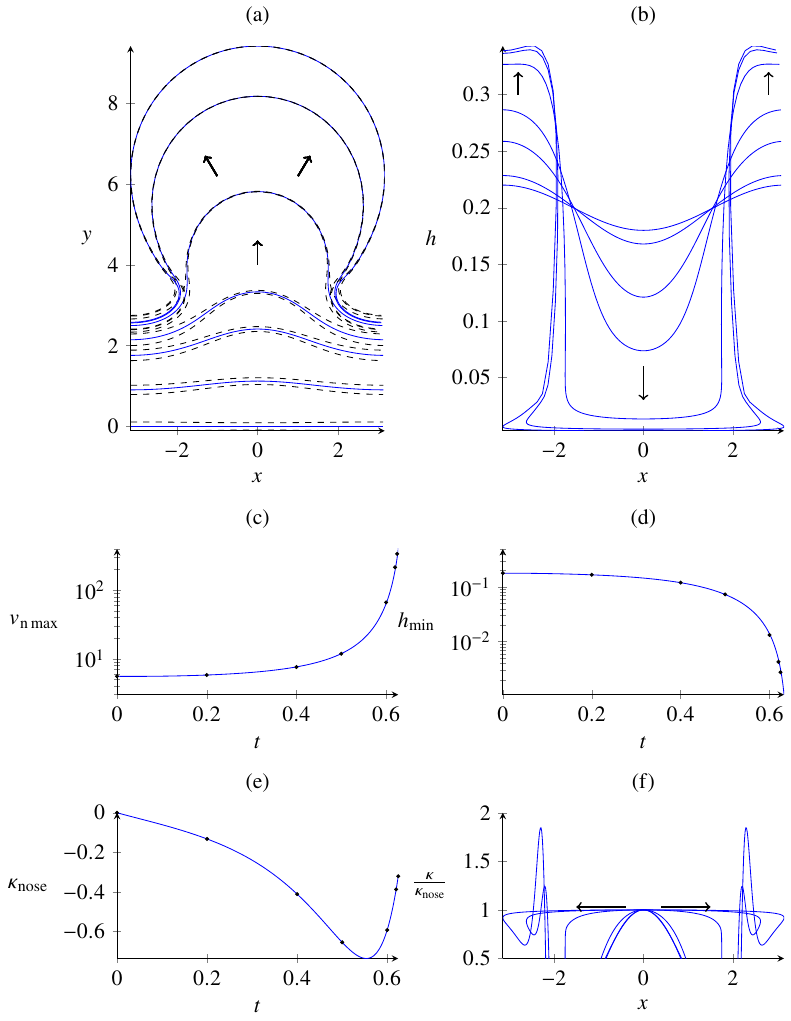}
\caption{Evolution of a filament with initial position $y = 0$ and thickness $h = 0.2(1-0.1\cos(x))$, with $P=1$ and surface tension $\sigma = 0.1$.  (a) the centreline profiles over time (solid) and the thickness (dashed lines), and (b) the thickness against $x$, show the initially thinner part of the filament bulge outward into a `bubble', while the bulk of the fluid is driven out to the edges of the filament.  Ultimately the profile intersects with the channel boundary at $x=\pm\pi$.  The (c) maximum normal velocity $v_\mathrm{n\,max}(t)$ and (d) minimum thickness $h_\mathrm{min}(t)$ appear to become unbounded and go to zero, respectively.  (e) The curvature at the nose ($x=0$) over time, initially increases in magnitude, then rapidly heads toward zero as the bubble expands.  (f) the curvature scaled by the curvature at the nose, showing the curvature tending to a constant over the bubble.  The dots marked in (c, d, e) correspond to the times at which profiles are plotted in (a, b, f), and the arrows in (a, b, f) indicate the direction of increasing time.}
\label{fig:sigma01}
\end{figure}

\begin{figure}
\centering
\includegraphics{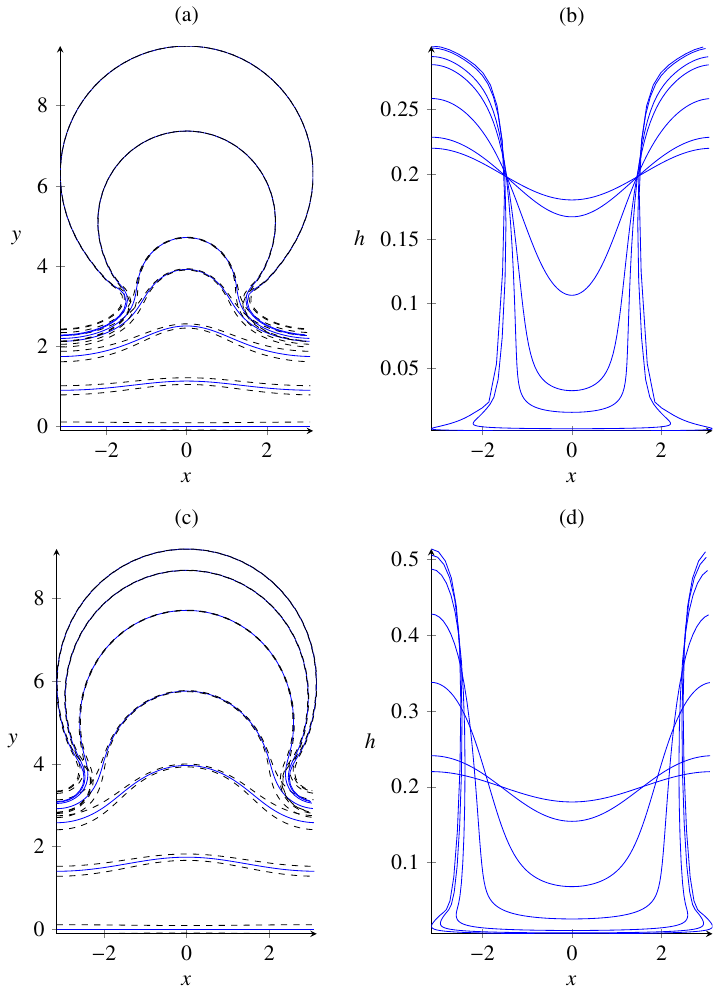}
\caption{Evolution of a filament with initial position $y = 0$ and thickness $h = 0.2(1-0.1\cos(x))$, for (a,b) $\sigma = 0.05$, and (c,d) $\sigma = 0.2$.  The formation of a circular bubble occurs in a similar fashion as happens for $\sigma = 0.1$, with the major effect being the width of the neck region that precedes the circular bubble.}  
\label{fig:otherSolutions}
\end{figure}

Here we use the front-tracking scheme to solve the thin filament model for later times, in the regime where the thickness becomes too small for the level set method to accurately resolve.  We use the initial condition
\[
y = 0, \qquad h = 0.2[1-0.1\cos(x)],
\]
which is the same as the initial condition of the exact solution \eqref{eq:exactsol1} depicted in Fig.~\ref{fig:exact_solutions}b.  In order to observe the effect of different surface tensions $\sigma$, we run simulations for $\sigma = 0.1$ (Fig.~\ref{fig:sigma01}), as well as $\sigma = 0.05$ and $\sigma = 0.2$ (Fig.~\ref{fig:otherSolutions}).

In the $\sigma=0.1$ simulation (Fig.~\ref{fig:sigma01}), the filament bulges outward in the centre near $x=0$, where the thin filament is initially thinner (and so the filament moves faster).  This bulge becomes a `bubble' that rapidly expands in radius while the thickness rapidly decreases.  The majority of the fluid is pushed into the outer regions of the channel.  At a finite time, the bubble intersects the channel walls at $x = \pm \pi$.  Unlike a fluid-filled Hele--Shaw cell, there is nothing to prevent the filament reaching the channel walls; this does not correspond to singular behaviour in the mathematical model, but physically represents an area of gas of lower pressure being trapped by the filament.

To further understand this behaviour, we note that \eqref{eqs:filamentEqs} has as a solution a perfectly circular bubble of radius $R(t)$, that evolves according to
\begin{equation}
R(t) = \frac{2\sigma}{P} + \left(R(0) - \frac{2\sigma}{P}\right) \e^{(P/c)t}, \qquad h = \frac{c}{R}.
\label{eq:bubble}
\end{equation}
Here $c$ is a constant that depends on the initial thickness.  In such a solution, the radius (if initially greater than $2\sigma/P$) grows exponentially, but does not exhibit a finite time singularity.  
It may be that this is the behaviour that is governing late stages of evolution of the filament depicted in this section.  In Fig.~\ref{fig:sigma01}e we plot the curvature $\kappa_\mathrm{nose}$ at the nose, or front, of the bubble (where $x=0$), while in Fig.~\ref{fig:sigma01}f we plot $\kappa/\kappa_\mathrm{nose}$.  The curvature initially grows in magnitude (in our convention the curvature is negative in the bubble) when the bulge initially grows, but then rapidly decreases in magnitude at the time when the bubble expands outwards.  When this happens, the curvature tends to a uniform state in the bubble region ($\kappa/\kappa_\mathrm{nose} \to 1$), implying the convergence to a circular shape.  While we have only plotted the interfaces up to the time at which it intersects the channel wall, mathematically the simulation continues for longer, and the bubble becomes more circular in shape.  Ultimately, the thickness becomes so small and the velocity so large that the numerical method no longer converges.

This behaviour is also seen for smaller and larger surface tension values $\sigma = 0.05$ and $0.2$, as depicted in Fig.~\ref{fig:otherSolutions}.  The notable effect of different surface tension is that the bulging region that forms the `neck' of the bubble is smaller or larger in width, as the surface tension is smaller or larger.

\subsection{Initial condition near a quasi-travelling wave}

In addition to the previous near-flat initial conditions, we demonstrate the instability of the quasi-travelling wave solutions considered in Section \ref{sec:travwave} by choosing an initial condition close to one such solution.
For simplicity we use the `grim reaper' exact solution for zero surface tension $K_0 = -\cos\theta$, and a scale factor $B=1$, for which
\[
y = \log(\cos x), \qquad h = \frac{b}{\cos x}
\]
(the exact solution is a good approximation for small $\sigma$, and although we do not show the results here, we have observed the same behaviour from numerically constructed initial conditions by solving \eqref{eq:travwaveK} for $\hat\sigma > 0$).  In terms of the quasi-travelling wave solution, the factor $b$ in the thickness $h$ is arbitrary as it corresponds to translation in time.  As the travelling wave is semi-infinite, we must of course approximate it by a finite curve that respects the required boundary conditions at $x=\pm\pi$.  We thus choose the closely related `hairclip' curve, defined by
\begin{equation}
y = \sinh^{-1}(\alpha\cos x), \qquad h = \frac{b}{\cos\theta} = b \sqrt{1 + \left(\frac{\alpha\sin x}{\cosh y}\right)^2}, \qquad \alpha \gg 1.
\label{eq:ICForTravellingWave}
\end{equation}
The larger the parameter $\alpha$ is made, the more elongated the initial finger becomes.  We note that the thickness will become large on the sides of the initial finger, representing the blow-up in thickness of the quasi-travelling wave in the limit $\theta\to\pi/2$ discussed in Section \ref{sec:travwave}.

\begin{figure}
\centering
\includegraphics{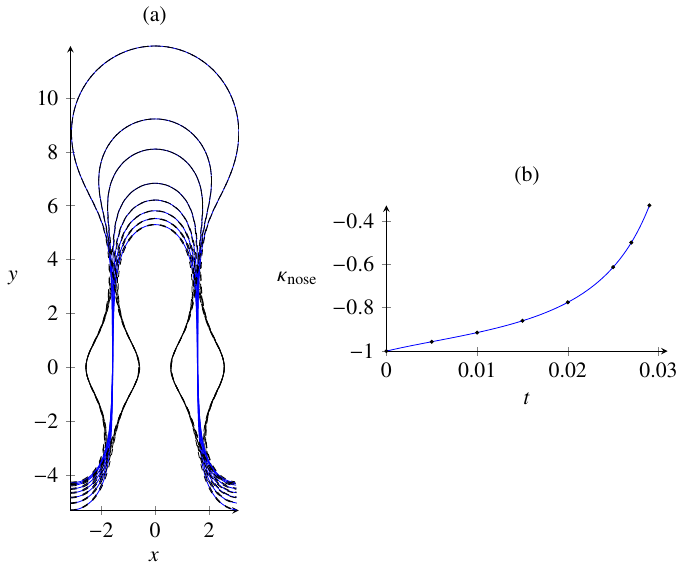}
\caption{
The evolution of a filament that starts near the ($\sigma=0$ limit) of the quasi-travelling wave described in Section~\ref{sec:travwave} (the precise initial condition is \eqref{eq:ICForTravellingWave}).  (a) the quasi-travelling wave appears unstable, with the front evolving toward the expanding circle seen for general initial conditions.  (b) The curvature at the nose of the filament ($x=0$) starts at $-1$ but decreases over time as the filament expands.  The dots marked in (b) correspond to the profiles plotted in (a).}
\label{fig:nearTravellingWave}
\end{figure}

We depict the result of such an initial condition in Fig.~\ref{fig:nearTravellingWave}, for the values $\sigma=0.05$, $\alpha=20$, $b=0.1$.  The regions with large thickness on the sides of the finger correspond to the exponential growth in thickness required for the travelling wave solution as $y\to -\infty$ (see Section~\ref{sec:travwave}); the filament in this region does not evolve to a great extent over the simulation.  At the front, while the finger does initially propagate forwards, the nose bulges outward and becomes more circular in shape, leading to the same late-time behaviour as seen for the more general initial conditions.

\section{Discussion}
In this paper we have developed a simplified but highly accurate second-order lubrication flow model \eqref{eqs:filamentEqs} that describes two-interface Hele--Shaw flow very well in regions where the thickness of the fluid region becomes small.  Due to the instability of one of the interfaces, this limit is one that is generally reached, even if initially the thickness is not very small.  By examining the generic singular behaviour of the leading-order model, even in the presence of surface tension (Section \ref{sec:leadingOrder}), we have shown why a second order-model is necessary to represent the dynamics of the full problem.  In particular, the fourth order spatial derivative term arising from the difference in curvature between the upper and lower interfaces is necessary to regularise the system.  Although we have included all terms formally of order $\epsilon^2$ in our model and simulations, we have also observed that the leading order model along with the addition of only this regularising term (proportional to $[hh_{sss}]_s$) will behave in a qualitatively similar manner (with small quantitative differences).  We have not shown these results here for the sake of brevity.

Here we note some clear differences between the instability of a thin filament, and the classical Saffman--Taylor instability in a semi-infinite fluid region that results in the Saffman--Taylor finger with (in the small surface tension limit) width half that of the channel.  One difference is that the thin filament model does not feel the effects of the channel wall away from the thick neck regions, and thus the rapidly expanding bubble may intersect the channel walls at finite time with no breakdown in the mathematical model.  Physically speaking, this phenomenon would correspond to trapping a part of the lower-pressure gas inside the fluid region.  In addition, as the walls have no strong effect, the orientation of the filament motion to be mainly in the positive $y$-direction is a somewhat artificial consequence of the initial condition.  For this reason it may be more natural to consider the fluid in an unbounded Hele--Shaw cell, with the length scale set by the initial thickness.

The specifics of the late stages of evolution of the filament (depicted in Section~\ref{sec:numerics}), wherein the thickness becomes very small and the velocity correspondingly large, is not resolved.  In order to understand the dynamics at late times of this system, an analysis of the stability of the quasi-travelling wave solutions, and the stability of the expanding circular bubble \eqref{eq:bubble} to non-radially symmetric perturbations, would be very valuable.
We observe that in our model, the filament does not appear to exhibit finite-time `bursting' behaviour, that is, the thickness does not go to zero at an isolated finite point in space and time.  In the case of self-similar breakup in the manner of the unforced lubrication equation described in \citet{Almgren1996b}, the thickness goes to zero at a point where the curvature becomes infinite, while in our solutions depicted in Section \ref{sec:numerics}, the thickness becomes small in the expanding circular region where the curvature is also decreasing, while the curvature is largest in magnitude in the neck regions, where the thickness does not decrease.
Bursting in physical systems is likely to require fully three dimensional effects to explain (that is, when the filament thickness becomes the same order as the separation between plates in the Hele--Shaw apparatus).  Once these two length scales are comparable, the filament model, and indeed the two dimensional Hele--Shaw model, are no longer valid.  The dynamics of the filament model studied in this work indicate that such a regime will generically be approached very rapidly in the form of the expanding bubble depicted in Section \ref{sec:numerics}.

\section*{Acknowledgements}The authors would like to thank Colin Please for many fruitful discussions.

\section*{Declaration of Interests}The authors report no conflict of interest.

\end{document}